# Recent development of magnetocaloric effect in Pyrochlore oxides

Vinod Kumar Dwivedi[1, *]

[1]*Department of Materials Science and Engineering,*
*Tel Aviv University, Tel Aviv 6997801, Israel*

## Abstract

In recent years, solid state magnetic cooling based on magnetocaloric effect (MCE) have drawn attention worldwide as a promising alternative potential candidate to the conventional gas compression-expansion cooling technique. In this chapter, the current developments of MCE in pyrochlore oxide materials is presented. The interaction between magnetic frustration, strong spin-orbit coupling, electronic correlations and crystal electric field lead to various complex magnetic states in pyrochlore oxides gives a broad template to explore these materials in MCE point of view. Pyrochlore oxides show strong frustration due to its crystal structure. Thus, the study of MCE in pyrochlore oxides could be fascinating not only from the an applications point of view but as well as from fundamental point of view. All reported litterateurs about MCE in pyrochlore oxides are $A_2Mn_2O_7$ (A = Er, Dy, Ho, Yb, La, Fe, Al, Ga, Sn), $Er_2Ti_2O_7$, $A_2TiMnO_7$ (A = Er, Dy, Ho), $A_2Mo_2O_7$ (A = Er, Dy, Gd, Y), $Pr_2Sn_2O_7$, $Er_2Ir_2O_7$ and $Y_2Ir_{2-x}Cr_xO_7$. Among all the compounds, cubic pyrochlore $Er_2Mn_2O_7$ exhibit largest conventional MCE value at change of field 5 T. On the other hand, $Y_2Ir_{2-x}Cr_xO_7$ series show coexistence of conventional MCE and inverse MCE which is absent in whole reported family of pyrochlore oxides. It is worth mention that depending on synthesis condition $A_2Mn_2O_7$ series crystallizes in either cubic pyrochlore phase ($Fd\bar{3}m$) or monoclinic layered perovskite (P2/M). It suggests the need of more materials to be explore experimentally and theoretically to establish the connection between these two phases.

---

[*]Electronic address: vinodd.iitbombay@gmail.com

# I. INTRODUCTION: THE MAGNETOCALORIC EFFECT

Now a days, in residential and commercial areas refrigeration technology has largest energy consumption rate i.e., 31% (USA), 40% (Europe) and 17% (across the globe) [1]. The demand for refrigeration technology is expected to substantially grow with the increase of population, improvement of living standard for the future. Therefore, the demand for environment friendly and energy-efficient cooling materials are needed to explore. Current cooling systems are based on the vapor compression-decompression mechanism uses hazardous greenhouse gases which impact badly on environment. Therefore, the search and development of environment friendly and energy-efficient cooling technology have drawn the attention of researchers worldwide. Among all available alternative refrigeration technologies such as thermo-ionic, thermo-electric, thermo-acoustic and magnetocaloric effect (MCE), researchers focused more on MCE based solid state cooling [2] because of its environment friendly and high efficiency natures. It has been proposed that refrigeration based on MCE mechanism may be the best alternative to the conventional gas compression cooling technology [3–15].

The MCE is defined as the change in temperature of a magnetic material when the magnetic field applied in an adiabatic condition ($\Delta T_{ad}$), and change in the magnetic entropy ($\Delta S_M$) when field is applied in an isothermal process. $\Delta S_M$ and $\Delta T_{ad}$ can be measured using isothermal magnetization data and indirectly from heat capacity data, respectively. The simple term, the conventional MCE is the alignment of localized magnetic moments when field is applied, leading to a reduction of magnetic entropy. Materials heat up with the application of magnetic field and cool down with the removal of field. However, the opposite behaviour (i.e., enhancement of magnetic entropy with the application of field) is also possible. Such effect is known as inverse magnetocaloric effect (IMCE) [16–20]. IMCE materials can be used as a heat sink in case of cooling with the application of magnetic field under adiabatic condition. Therefore, the search for IMCE materials is important in the discovery of potential conventional MCE materials for development of solid-state cooling. Moreover, the materials exhibiting MCE near room temperature is also equally important and have been reported in various materials [21–25].

However, the search for suitable MCE materials is going on almost from a century, no satisfactory materials based on MCE is found for practical application so far. The hindrance in the development of solid-state magnetic refrigeration technology based on MCE are expensive magnetic coolant materials, narrow working window and requirement of high magnetic field. The thorough details can be found in the various review papers based on the MCE and IMCE listed in the current review article. Therefore, the discovery of new MCE materials with suitable magnetic cooling parameters are equally important. For this purpose, the current review focuses especially on the strongly frustrated classical Heisenberg antiferromagnets pyrochlore oxides (It has been proposed theoretically to host a larger field-induced adiabatic temperature change as compared to non-frustrated magnets [26, 27]) which is less explored till date.

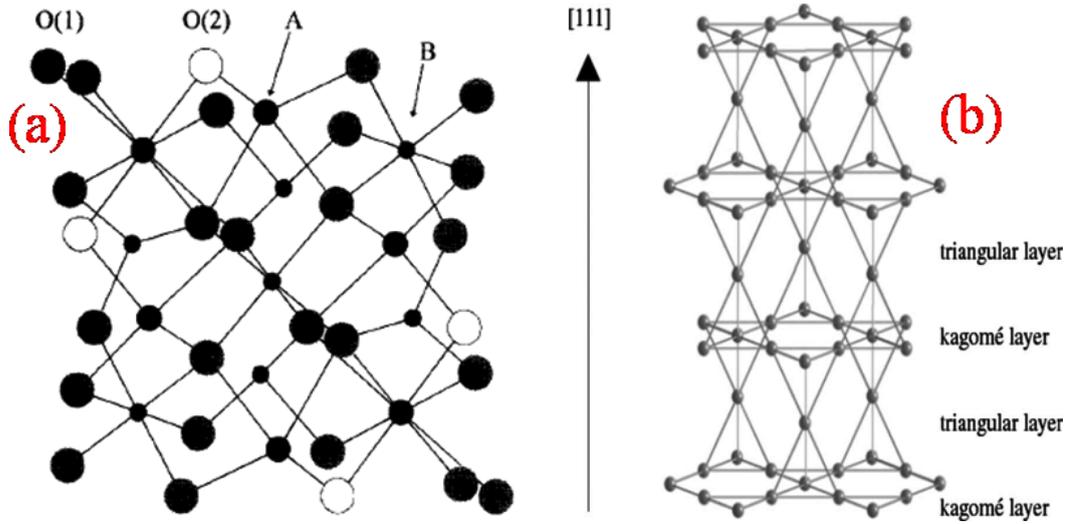

FIG. 1: (a) View of the $A_2B_2O1_6O2$ pyrochlore structure. (b) Alternating kagome and triangular planar layers stacked along a [111] direction of the pyrochlore lattice. All pyrochlore iridates show trigonal compression of the oxygen cages along the [111] directions. Adapted from [28, 29].

## II. PYROCHLORE OXIDES: - STRUCTURE AND MAGNETIC FRUSTRATION

This section will give an outlook about the magnetic frustration arising due to geometry of pyrochlore structure, and how it plays a crucial role in determining the physical properties and ground state of such materials.

### A. Crystal structure of pyrochlore oxide

Pyrochlore materials get their name from the mineral $NaCaNb_2O_6F$, and the structure of pyrochlore was first reported by von Gaertner in 1930. Currently, the general formula $A_2B_2O_7$ or $A_2B_2O1_6O2$ for pyrochlore oxide is in standard usage. Here, A is usually trivalent rare earth ion or Y, and B is usually a transition metal ion. It should be mentioned that it is possible to form $A_2^{2+}B_2^{5+}O_7$ (2+, 5+) and $A_2^{3+}B_2^{4+}O_7$ (3+, 4+). However, the former combination (2+, 5+) are not common and have not been studied in detail. The structure of (3+, 4+) combination pyrochlore oxide exhibit the $A^{3+}$-cation at the Wyckoff position 16d-site with minimal coordinates (0.5, 0.5, 0.5), $B^{4+}$-cation at 16c-site as an origin (0, 0, 0), oxygen anions $O1^{2-}$ at 48f-site (p, 1/8, 1/8) and $O2^{2-}$ at 8b (3/8, 3/8, 3/8) with space group $Fd\bar{3}m$ (227). Here, p is only one adjustable positional parameter.

**Local structure of the A and B sites** - The pyrochlore structure $A_2B_2O_7$ made of with two types of cations. The larger cation A and smaller cations B are eight and six coordinated, respectively. In $A_2B_2O1_6O2_1$ structure the coordination arrangement about A and B sites is governed by the value of positional parameter p. A perfect octahedron can be found about the B(16c)-site for p = 0.3125 and for p = 0.375 a ideal cube about A(16d)-site. Usually, for most of the materials the value of positional parameter is found in the range of p = 0.320 - 0.345. The smaller B-type cations are bonded to six O1 atoms at equal distances, i.e. B-O1 bond length must be of equal length as shown in Fig. 1a. The O1-B-O1 bond angles vary between $81^0$-$100^0$, with slight deviation from the ideal octahedral value of $90^0$. Although, the angle B-O1-B, in which the O1 atom is shared between two octahedra is restricted to a very narrow range in pyrochlores. Usually between $127^0$ to $134^0$, the distortion of the A-site geometry from a perfect cube is very large. Generally, the larger A cations are coordinated by six O1 atoms and two O2 atoms making an axially compressed scalenohedron as depicted in Fig. 1a. The A-O1 and A-O2 bond distances show a large difference. While the typical values of A-O1 are 2.4-2.5 Å, the A-O2 bonds have shortest ($\approx$ 2.2 Å) bond length best known for any rare-earth oxide. Therefore, the A-site has very prominent axial symmetry along a local < 111 > direction. Moreover, the shorter A-O2 bond length depends only on lattice parameter a, the A-O1 and B-O2 bond distances depend on both the lattice parameter and the variable positional parameter p.

An alternative visualization of pyrochlore structure shows stacked alternating kagome and triangular planar layers of A and B-sites along [111] direction as shown in Fig. 1b. Henceforth, the pyrochlore structure can be seen as consisting of two interpenetrating networks of $B_2O1_6$ [in which each of the oxygen (O1) atoms of $BO1_6$ octahedra are corner shared and can be seen as forming regular $B_4$ tetrahedra] and $A_2O2$ (each oxygen O2 is coordinated with four A cations to form a second tetrahedral network) sublattices. The $B_2O_6$ network is fairly rigid and does not interact strongly with the $A_2O$ network and as a result cation and/or anion vacancies in the $A_2O$ network do not significantly reduce the stability of the lattice. Consequently, in the oxygen deficient pyrochlores, while anion vacancies can occur at either site, partial loss of the O2-type anions appear to be favoured [28, 29].

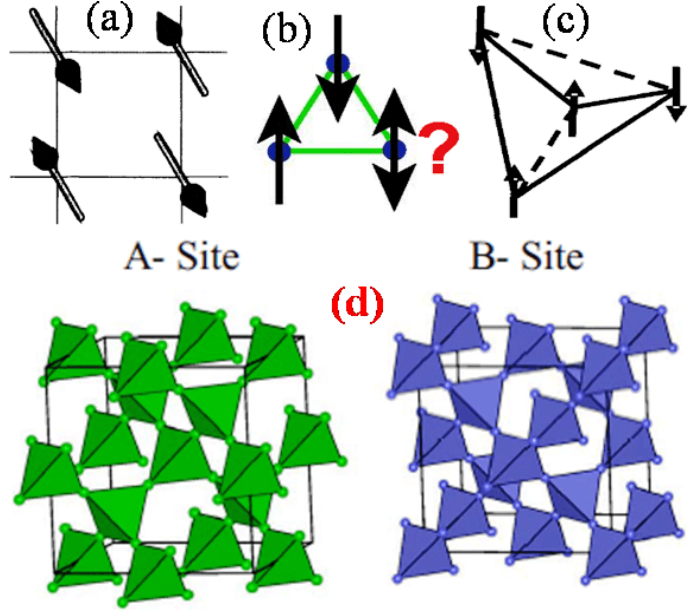

FIG. 2: Antiferromagnetic nearest neighbour interactions on the (a) square lattice, (b) triangular lattice and (c) tetrahedron. Ising spin on a tetrahedron showing four satisfied (antiferromagnetic) and two frustrated ferromagnetic bonds (dashed line). (d) The *A* and *B* sublattices of the cubic pyrochlore materials. Either or both sublattices can be magnetic. Adapted from [29, 30].

### B. Magnetic frustration

Competing interactions or frustration are common in condensed matter systems. A system is said to be frustrated whenever it cannot minimize its total classical energy by minimizing the interaction energy between each pair of interacting degrees of freedom, pair by pair [29–31]. Figure 2a shows spins reside on the square lattice easily satisfy the requirement by minimizing its total classical energy leading to absence of frustration. Frustrated magnetic systems can be classified in two types. The first type is because of the lattice geometry and the second in which site or bond randomness is a key factor. The simplest case of geometric frustration is that of Ising spins which can only point in either up or down directions, interacting via nearest neighbour antiferromagnetic exchange interaction and lie at the vertices of an equilateral triangle shown in Fig. 2b. The tetrahedron shown in Fig. 2c is three-dimensional version of a scenario similar to triangular lattice with four near neighbour spins situated at the vertices of tetrahedron. The triangular lattice made of many edge-sharing triangles, has very large degeneracy or configurational spin disorder, which leads not only to a zero-temperature residual entropy but absence of symmetry breaking phase transition down to absolute zero temperature. This is an example of geometric frustration where the regular periodic configuration (triangular, kagome, tetrahedral or otherwise) of the space lattice prevent to develop a long range ordered state. The same system containing uniaxial Ising spins with ferromagnetic interaction does not show frustration.

The second type of frustration comes into picture due to 'quenched randomness'. The randomness in systems can be introduced by many ways. Other than impurities by doping, emergence of vacancies, grain boundaries or dislocations in a crystal give rise to disorder in a system. Besides reducing long range ordering temperature, the quenched or frozen-in disorder systems where defects or impurity atoms do not move over considerable time scales could also lead to non-trivial magnetic phases such as spin glass, Griffiths phase [31, 32], etc. The dilution of magnetic moments (or substitution of non-magnetic atoms) in a disorder free geometrically frustrated system can also generate random frustration. A spin glass system is defined as a random mixed-interacting magnetic system characterized by a transition from a paramagnetic state of thermally fluctuating spins to a glass like state of spins frozen in time but random in direction below a certain temperature called freezing temperature [31]. In contrast to pure magnets (disorder free), long range magnetic ordering is absent in spin glass like systems. The most common examples of spin glass systems are $Cu_{1-x}Mn_x$ and AuFe (magnetic iron atoms in a gold matrix), in which a very little amount of magnetic impurity (x $\sim$ 1%) is introduced in non-magnetic host Cu and Au, respectively [31]. Experimentally the level of frustration in magnetic system is measured by frustration index f and defined as [30]

$$f = \frac{|\theta_{CW}|}{T^*} \quad (1)$$

where, $\theta_{CW}$ is the Curie-Weiss temperature, and can be obtained from the Curie-Weiss fit to the high temperature paramagnetic part of the inverse dc susceptibility.

$$\chi = \frac{C}{T - T^*} \quad (2)$$

$T^*$ would be the freezing temperature $T_f$ for spin glass system. Heavily frustrated system has

very low T* as compared to $\theta_{CW}$.

The geometrical frustration is closely connected to the pyrochlore oxides. In the cubic pyrochlore oxides, $A_2B_2O_7$ (A = trivalent rare earth ion which includes the lanthanides, Y, and sometimes Sc Y, Bi, B = either a transition metal or a p-block metal ion), the A and B ions reside on two different interpenetrating lattices of corner-sharing tetrahedra as shown in Fig. 2d. In $A_2B_2O_7$, if A, B, or both ions are magnetic and exchange interaction between nearest neighbour is antiferromagnetic, then the system can be highly frustrated by geometry. Therefore, pyrochlore oxides exhibiting antiferromagnetically coupled spins, may not develop a long-range magnetic ordering. Examples of such phenomena exhibited by pyrochlore oxides are spin glass in $Y_2Mo_2O_7$ [33], spin liquid behaviour in $Tb_2Ti_2O_7$ [34], spin-ice phenomena [35] in $Ho_2Ti_2O_7$, $Dy_2Ti_2O_7$ and $Tb_2Sn_2O_7$ [36], unconventional anomalous Hall effect in metallic $Nd_2Mo_2O_7$ [37, 38], superconductivity in $Cd_2Re_2O_7$ [39] and the possible frustrated Kondo-lattice like effect in $Pr_2Ir_2O_7$ [40, 41], etc. These few examples show the importance of geometrical frustration in pyrochlore oxides.

**C. Estimation of important MCE parameters**

The process of calculation for the few important parameters are given below.

(I) The change in magnetic entropy $\Delta S_M$ using Maxwell relation [7] from the M-H isotherms

$$\Delta S_M = \int_0^H (\partial M/\partial T)\, dH \quad (3)$$

(II) The relative cooling power (RCP) which is often used to estimate their potential for magnetic cooling

$$RCP = -\Delta S_M^{max} \delta T_{FWHM} \quad (4)$$

(III) The refrigerant capacity (RC) gives an estimate of the transferred amount of heat between the hot end at $T_{hot}$ and cold end at $T_{cold}$. It is calculated as the area under the $-\Delta S_M(T)$ curve between $T_{hot}$ and $T_{cold}$ for a particular magnetic field [as shown in Fig. 3] i.e.,

$$RC = \int_{cold}^{hot} [\Delta S_M(T)]\, dT \quad (5)$$

The two temperatures $T_{cold}$ and $T_{hot}$ are associated to the working range of the refrigerator, which is related to the full width at half maximum ($\delta T_{FWHM}$) of $-\Delta S_M(T)$ curve. The mate-rials exhibiting larger RC is preferred for transfer the larger amount of heat while comparing the different materials to be utilized in the same thermodynamic cycle.

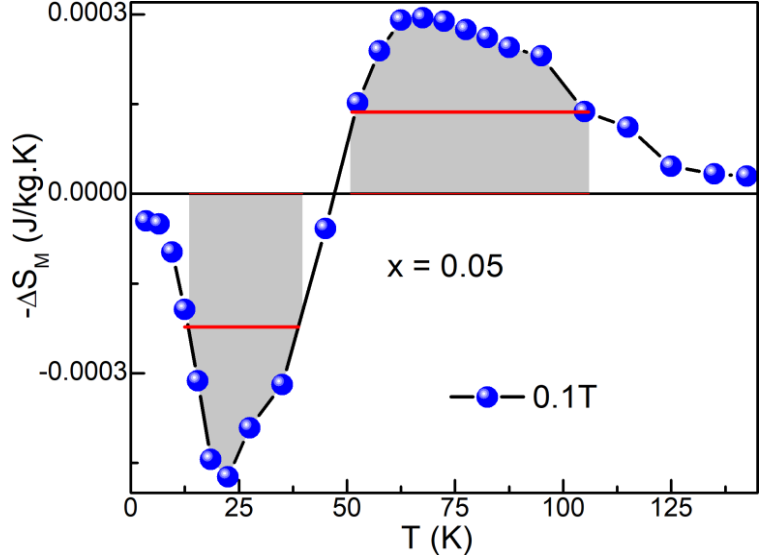

FIG. 3: Calculation method for RC and RCP from the $\Delta S_M$(T) curves.

## III.  MCE IN PYROCHLORE OXIDES $A_2B_2O_7$

Long back, theoretically it was proposed that the strongly frustrated classical Heisenberg antiferromagnets could host large adiabatic temperature change induced by applied field than nonfrustrated magnetic systems [26, 27]. Pyrochlores are known to exhibit strong magnetic frustration. It have drawn the attention of researchers worldwide to investigate the MCE in such highly frustrated pyrochlore systems. The following section discuss the progress of MCE in various pyrochlore oxide systems.

### A.  Mn based pyrochlore oxides

The first MCE investigation was done on the compound $Er_2Mn_2O_7$ [42]. In this material, 4f-localized magnetic moments of $Er^{3+}$ and 3d-metal electrons of $Mn^{4+}$ both are magnetic giving rise to a competition between the magnetic interactions. Synthesis of polycrystalline $Er_2Mn_2O_7$ have performed by solid-state reaction method heated at 1300 $^0$C for 48 h in air using the ingredi- ents $Er_2O_3$ and $MnO_2$ reported elsewhere [42]. The refinement of X-ray diffraction (XRD) pattern suggests monoclinic phase with space group P2/M. The magnetization (M) as a function of temperature (T) measured at different applied magnetic field enhances with field [Fig. 4a] suggest the presence of ferromagnetic components. The magnetic transition at Neel temperature $T_N$ estimated from minimum of dM/dT versus T curve shown in inset of Fig. 4a. The values of $T_N$ are 1.93 K (0 T), 1.97 K (0.05 T), 2.45 K (1 T), 3.25 K (2 T), 4.25 K (3 T), 5.01 K (4 T) and 5.47 K (5 T). The Curie-Weiss temperature $\theta_{CW}$ is calculated by fitting the $(\chi - \chi_0)^{-1}$ versus T curve using Eq. 2 and shown in Fig. 4b. The estimated values of $\theta_{CW}$ are -24.05 K (0.05 T), -16.55 K (1 T), -15.16 K (2 T), -14.47 K (3 T), -14.23 K (4 T) and -13.92 K (5 T). The negative values of $\theta_{CW}$ indicates the presence of antiferromagnetic (AFM) exchange interactions. It is obvious that the value of $T_N$ and $\theta_{CW}$ varies with applied magnetic field. Figure 4c shows the magnetization isotherms as a function of field measured at various temperature around $T_N$. Although, M-H does not saturate upto 10 T applied magnetic field, the saturation at low fields (1-3 T) and the sharp rise in M near $T_N$ indicates the presence of ferromagnetic interactions. The change in magnetic entropy $\Delta S_M$ as a function of T measured at several field change 0-5 T is shown in Fig. 4d. The maximum value of $\Delta S_M$ is found to be 1.6 J/K-mol ($Er_2Mn_2O_7$) at field change of 10 T. It shows a maximum around $T_N$ shifts towards higher temperature when magnetic field increases.

Usually, for a system $A_2B_2O_7$, if the cation ionic ratio [IRR = $r(A^{3+})/r(B^{4+})$] lie between 1.46-1.80, a pyrochlore crystal structure is expected at atmospheric pressure [29,43,44]. However, the value of IRR can be extended upto 2.3 for the samples synthesized at high pressure and temperature [43]. Moreover, for the value of IRR = 1.97 to 2.24, a pseudo-pyrochlore structure is also expected [44]. For $Er_2Mn_2O_7$ compound, IRR (= 1.894) falls out of the pyrochlore range. It favors that $Er_2Mn_2O_7$ must form in monoclinic layered perovskite structure when synthesized at ambient pressure. Therefore, for smaller B-cations like $Mn^{4+}$, the cubic pyrochlore structure may only be synthesized by high-pressure techniques.

In order to prepare the pyrochlore cubic phase with space group $Fd\bar{3}m$, polycrystalline $Er_2Mn_2O_7$ have synthesized using high pressure and high temperature methods. The details of synthesis and characterizations can be found elsewhere [45]. The cubic pyrochlore crystal structure has been confirmed using XRD and neutron powder diffraction (NPD) techniques. Figure 5a reveals the magnetic susceptibility ($\chi$) as a function of T measured at applied magnetic field $\mu_0H$ = 0.1 T following the protocol zero field cooled warming (ZFCW), field cooled warming (FCW) and field cooled cooling (FCC). The three protocols are defined as; (I) ZFCW protocol - sample cooled in absence of any magnetic field from room temperature down to lowest desired tempera-ture and data were recorded during warming up condition, (II) FCW protocol - sample cooled in the presence of magnetic field from room temperature down to lowest temperature and data were recorded during warming up condition, and (III) FCC protocol - sample were cooled in the presence of magnetic field from room temperature down to lowest temperature and data were recorded immediately as cooling process started. It is obvious from Figure 5a that $\chi$ versus T curve shows sharp rise with lowering the temperature suggests FM transition at $T_C = 34.5$ K. A bifurcation can be seen between ZFC and FC curve at $\sim$15 K. There is no splitting observed between FCw and FCC curves indicating absence of thermal hysteresis in accordance with the second order magnetic phase transition (SOMPT). The inverse magnetic susceptibility $\chi^{-1}$ versus T curve is fitted at high temperature ranging 50-300 K by Eq. 2 shown in the inset of Fig. 5a. The value of $\theta_{CW}$ is found to be +37.4 K. The positive value of $\theta_{CW}$ suggests a dominant FM exchange interaction between the spins. Figure 5b shows field dependent Magnetization measured at 2 K between -7 to +7 T field suggests FM nature of $Er_2Mn_2O_7$ compound with huge value of saturation magnetic moment 18.9 $\mu_B$/f.u. (189.6 emu/g). Inset of Fig. 5b demonstrates a small hysteresis with coercive field $\mu_0H_C \approx 0.02$ T at 2 K indicates presence of weak magnetic hysteresis. Figure 5c reveals $\Delta S_M$ versus T curves spans over a large temperature range centered at $T_C$ measured at various field change 0-7 T. The $\Delta S_M$ attains maximum value 2.86, 5.3, 8.7, 11.4, 13.8, 16.1, 18.0, and 20.1 J/K-kg at various field change 0.05 T, 1 T, 2 T, 3 T, 4 T, 5 T, 6 T, and 7 T, respectively shown in Fig. 5d. It is clear from Fig. 5d, that the field dependent RC (RCP) increases linearly and reaches to the value 68 (90), 288 (377), 522 (680), and 747 J/kg (964 J/kg) under magnetic field change of 1, 3, 5, and 7 T, respectively. It is visible that the value of RC is quite large for $Er_2Mn_2O_7$ compound. Such a giant RC is associated to the broad $\Delta S_M$-T curves imply the wide working temperature window ranging 20-80 K. In addition, $Er_2Mn_2O_7$ possess a large $\Delta S_M$ = 5.3 J/K-kg and RC = 68 J/kg values at low field change of 1 T, makes it a potential low field MCE candidate suitable for the permanent magnets.

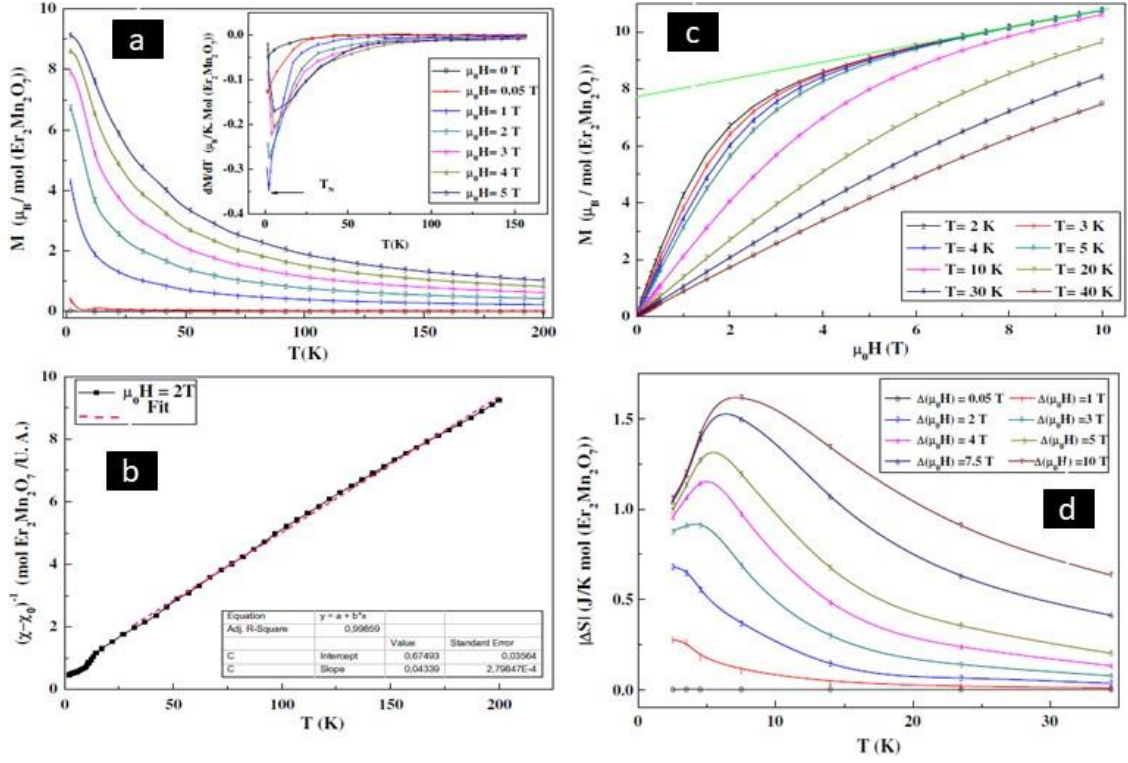

FIG. 4: (a) Magnetization (M) versus temperature (T) measured at various field. Inset: $\frac{dM}{dT}$ as a function of temperature at different applied magnetic field. (b) inverse magnetic susceptibility $(\chi - \chi_0)^{-1}$ versus T of monoclinic $Er_2Mn_2O_7$ sample at $\mu_0H = 2$ T. (c) Magnetic isotherms as a function of field measured in a temperature ranging of 2-40 K. (d) isothermal magnetic entropy change $\Delta S_M$ (T) versus T performed at different magnetic fields. Adapted from [42].

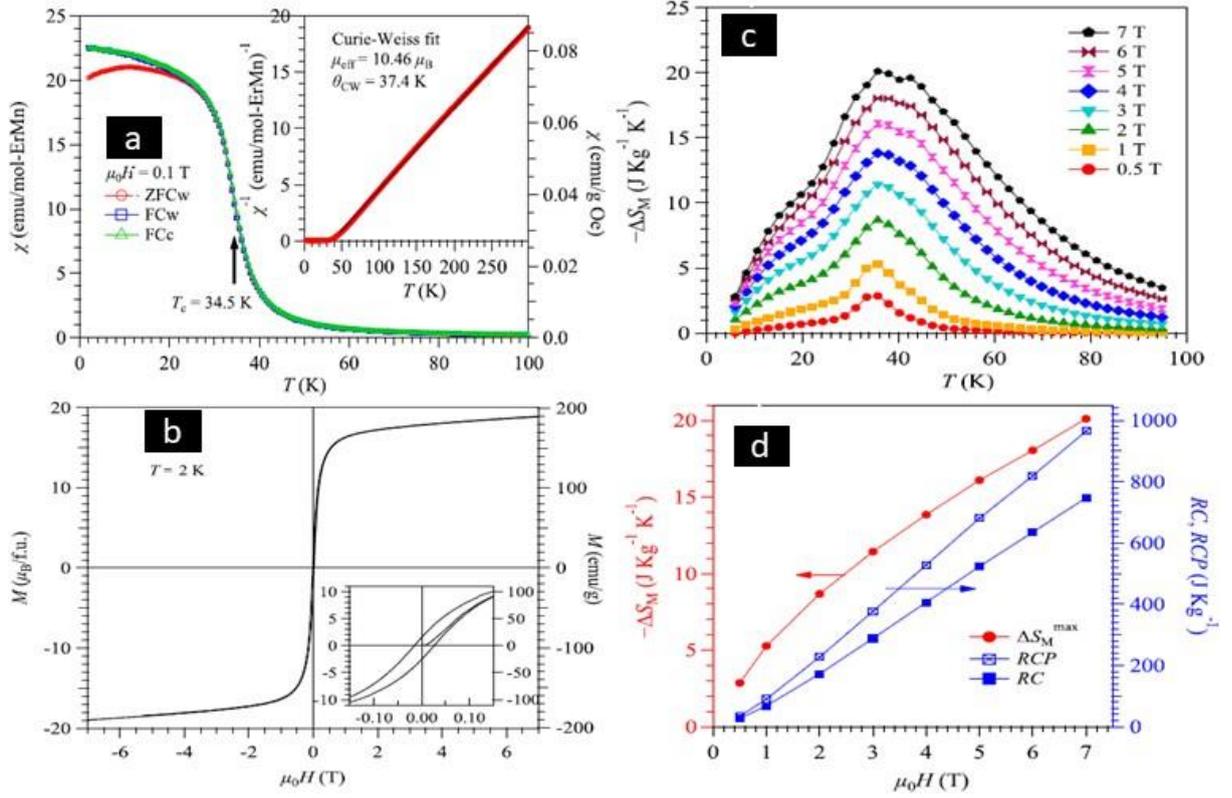

FIG. 5: (a) DC magnetic susceptibility $\chi$ as a function T for cubic pyrochlore $Er_2Mn_2O_7$ compound measured at applied magnetic field $\mu_0H = 0.1$ T following the protocol ZFCW, FCw, and FCc. The ferromagnetic transition temperature $T_C$ = 34.5 K was estimated from the minimum of $\frac{d\chi}{dT}$. Inset shows the Curie-Weiss law fitted inverse $\chi^{-1}$(T) curve. (b) Magnetization as a function of $\mu_0H$ measured at 2 K. Inset; enlarged view of the M($\mu_0H$) curve. (c) $\Delta S_M$ (T) versus T measured at various field changes upto 7 T. (d) variation of the maximum $\Delta S_M$ (T), the refrigerant capacity (RC), and the relative cooling power (RCP) as a function of applied magnetic field. Adapted from [45].

In pyrochlore $Er_2Mn_2O_7$, $Mn^{4+}$ ($3d^3$, S = 3/2) sublattices order ferromagnetically due to Mn-O-Mn bond angle 130$^0$ (favors FM alignment). It reduces the geometrical frustration and produces a strong internal field to polarize the $Er^{3+}$ ($4f^{11}$, S = 3/2, L = 6, J = 15/2) sublattices, which leads to a two-sublattice cooperative FM alignment of spins giving rise to large reversible MCE at $T_C$. It is obvious that two sublattice cooperative FM ordering plays an important role for the giant MCE values in cubic pyrochlore oxide $Er_2Mn_2O_7$ with space group $Fd\bar{3}m$ synthesized at high pressure high temperature. While it is absent in the isomeric $Er_2Mn_2O_7$ compound crystallizes in a monoclinic layered perovskite structure with P2/M space group synthesized at ambient pressure [42]. Pyrochlore $Er_2Mn_2O_7$ compound shows FM transition at $T_C$ = 34.5 K, while monoclinic layered perovskite $Er_2Mn_2O_7$ goes under AFM transition at $T_N$ = 2 K.

Further the MCE of other members of FM pyrochlore oxides $A_2Mn_2O_7$ (A = Dy, Ho, Yb) series have also been investigated [46]. Here, $Dy^{3+}$, $Ho^{3+}$, and $Yb^{3+}$ have larger magnetic moment than $Er^{3+}$. The $A_2Mn_2O_7$ cubic pyrochlore series can be prepared under high-pressure high temperature conditions. Whole series show a cooperative SOMPT FM transition at $T_C \approx$ 39 K [Fig. 6]. Similar to $Er_2Mn_2O_7$ cubic pyrochlore oxide, $A_2Mn_2O_7$ (R = Dy, Ho, Yb) series dis- play reversible and large value of MCE around $T_C$. The maximum value of $\Delta S_M$ (RC) under 7 T magnetic field change for $A_2Mn_2O_7$ series reaches to 20.1 J/K-kg (564 J/kg) for $Dy_2Mn_2O_7$, 18.4 J/K-kg (503.7 J/kg) for $Ho_2Mn_2O_7$ and 13.9 J/K-kg (371.4 J/kg) for $Yb_2Mn_2O_7$. The series possess lesser values than expectation (larger magnetic moments of $Dy^{3+}$, $Ho^{3+}$, and $Yb^{3+}$ ion compared to $Er^{3+}$). In pyrochlore $Er_2Mn_2O_7$, MCE phenomena associated to two-sublattice co- operative FM ordering, while for $A_2Mn_2O_7$ (R = Dy, Ho, Yb) series, strong single-ion anisotropy prevents the complete FM ordering of magnetic moments of $A^{3+}$ ion.

In a recent study, the effect of ionic radius on the evolution of MCE and Griffiths phase have been performed in new type of compound $A_2Mn_2O_7$ (A = La, Fe, Al) [47–50]. The series of materials were synthesized by conventional solid-state route at ambient pressure. The refinement of XRD pattern suggests monoclinic layered perovskite crystal structure with P2/M space group.

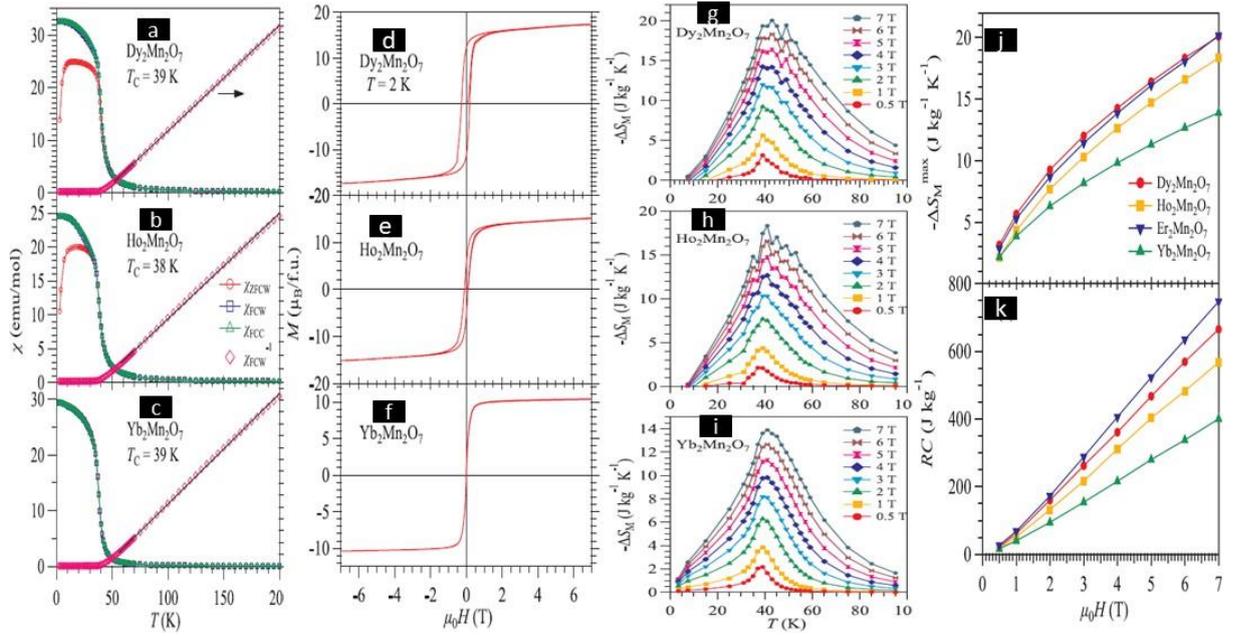

FIG. 6: Temperature dependent dc magnetic susceptibility $\chi$ and $\chi^{-1}$ of (a) $Dy_2Mn_2O_7$, (b) $Ho_2Mn_2O_7$, and (c) $Yb_2Mn_2O_7$ compound measured at H = 0.1 T following the protocol ZFCw, FCw and FCc. Curie- Weiss fitting of $\chi^{-1}$-T curve in paramagnetic range is represented by Solid lines. M as a function of H measured at 2 K for (d) $Dy_2Mn_2O_7$, (e) $Ho_2Mn_2O_7$, and (f) $Yb_2Mn_2O_7$. Temperature dependence of $\Delta S_M$ curves measured at various field changes upto 7 T for (g) $Dy_2Mn_2O_7$, (h) $Ho_2Mn_2O_7$, and (i) $Yb_2Mn_2O_7$. (j) the maximum of $\Delta S_M^{max}$ and (k) RC as a function of field. Adapted from [46].

All three samples $La_2Mn_2O_7$, $Fe_2Mn_2O_7$, and $Al_2Mn_2O_7$ show a FM transition and Griffiths-like state at low temperature [Fig. 7]. Figure 7d display $\Delta S_M$-T curves for 5 T filed change. It is obvious that the value of $\Delta S_M$, RCP and strength of Griffiths-like phase enhances by reducing the ionic radius of $A^{3+}$-site.

Very recently the magnetic and MCE properties have been explored in $Ga_2Mn_{2-x}Cr_xO_7$ (x = 0.0, 0.1, 0.3, and 0.5) series [51]. Samples were synthesized by solid state reaction route at ambient pressure and found to be crystallize in cubic pyrochlore phase with space group $Fd\bar{3}m$. All samples display a AFM transition at $T_N \sim$ 42 K along with glassy behaviour around 12 K and Griffiths-like state. Figure 8 shows $\Delta S_M$-T curve for $Ga_2Mn_{2-x}Cr_xO_7$ series. The value of -$\Delta S_M$ and RCP are 1.36 J/K-kg and 42.96 J/kg (x = 0.0), and 1.47 J/K-kg and 47.73 J/kg (x = 0.5). It suggests a slight increment in the value of -$\Delta S_M$ and RCP with Cr doping likely due to the presence of mixed oxidation states of Mn i.e., $Mn^{4+}$ and $Mn^{3+}$. A minimum can also be seen in the -$\Delta S_M$-T curve at

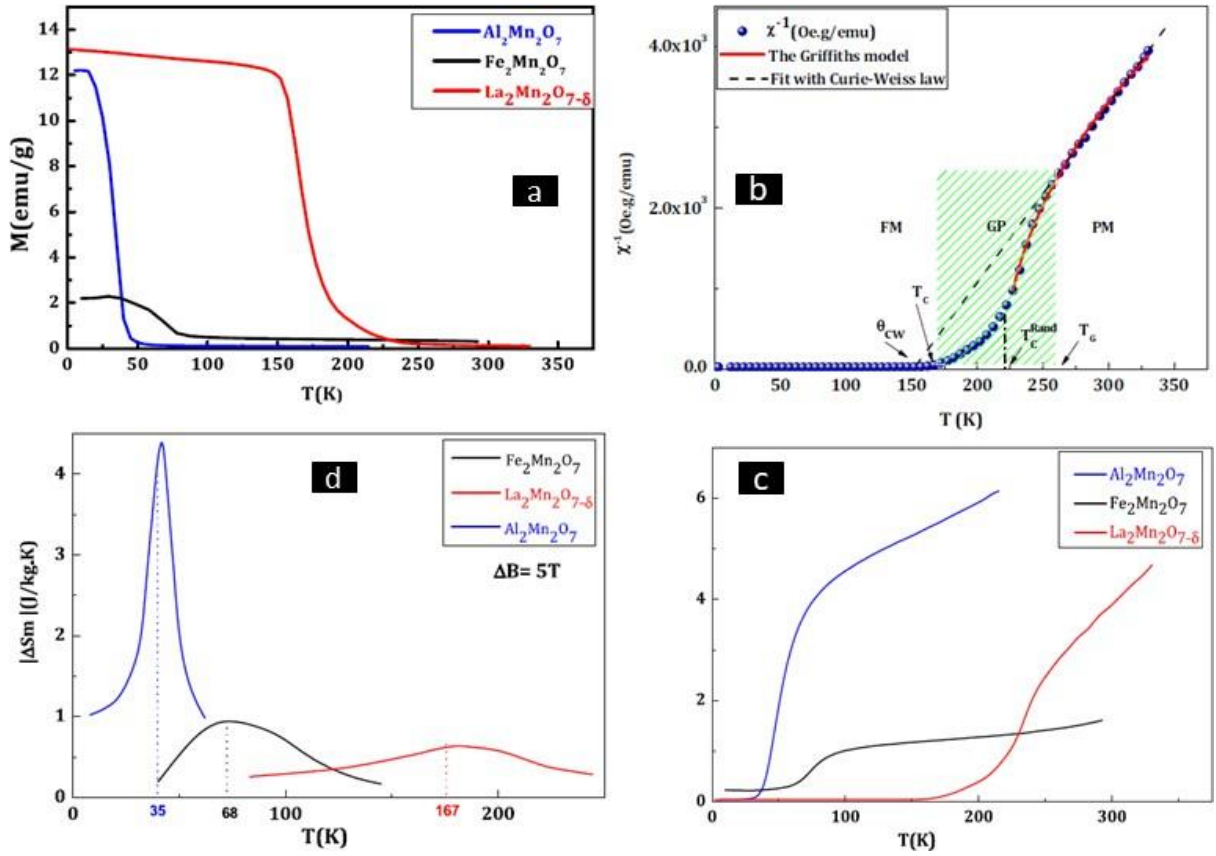

FIG. 7: (a) Temperature dependent dc magnetization, (b) $\chi^{-1}$ versus T, (c) $\chi^{-1}$ as a function of T and (d) $\Delta S_M$-T curves measured at field changes upto 5 T for the series compounds $A_2Mn_2O_7$ (A = La, Fe, Al). Adapted from [47–50].

∼10 K emerging due to spin reorientation. In other recent paper reported by the same group [52], magnetic and MCE properties has been explored in a new pyrochlore compound $Sn_2Mn_2O_7$. It has synthesized by conventional solid-state route at atmospheric pressure. Refinement of XRD pattern suggest monoclinic layered perovskite phase with space group P2/M. The XPS study shows 4+ and 3+ valence state for Sn and Mn, respectively. It undergoes a second order FM transition at $T_C$ = 41 K. The estimated value of -$\Delta S_M$ and RCP are found to be 2.94 J/K-kg and 41 J/kg, respectively for the field change of 5 T. It can be notice that the value of $\Delta S_M$ is slightly higher than what is reported for $Fe_2Mn_2O_7$ and less than $Al_2Mn_2O_7$.

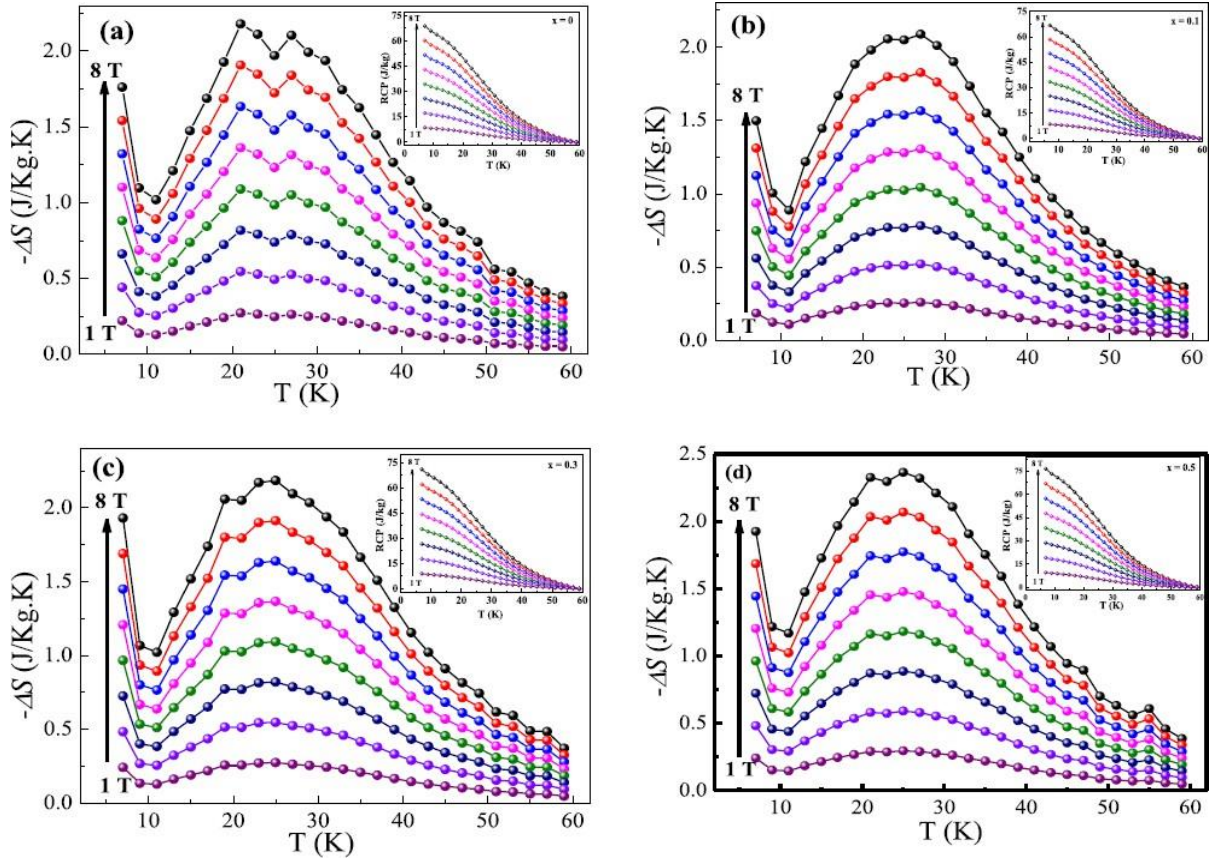

FIG. 8: $\Delta S_M$-T curves measured at field changes upto 8 T for the series $Ga_2Mn_{2-x}Cr_xO_7$ (a) x = 0.0, (b) x = 0.1, (c) x = 0.3 and (d) x = 0.5. Inset shows RCP as a function of temperature. Adapted from [51].

**B. Ti based pyrochlore oxides**

The Pyrochlore oxide $Dy_2Ti_2O_7$ has been known for its spin ice behaviour at low temperature. This compound obeys the ice rule i.e., in each basic tetrahedron, two spins point outward and two spins point inward (2-in/2-out). Experimentally, $Dy_2Ti_2O_7$ shows no phase transition at zero magnetic field down to 50 mK. The MCE properties has been explored in $Dy_2Ti_2O_7$ single crystal. Synthesis and characterization can be found elsewhere [53]. Field dependence of $\Delta S_M$ has been measured in a magnetic field H ∥ [111]. The estimated released entropy across the transition found to be 0.5 J/(K-mol-Dy). The peak intensity of S(H) depends strongly on temperature. The peak at T = 400, 500 mK is very prominent at the spin-flip field ∼ 0.9 T, attributed to the two-in, two-out and the three-in, one-out states. The MCE in Y doped $Dy_2Ti_2O_7$ samples have also been examined at 0.3 K to 6 K in applied field up to 2 T. Polycrystalline powdered samples have synthesized by well-established solid state reaction method following the protocol reported elsewhere [54]. It has proposed that in the spin ice regime slow relaxation process may be studied qualitatively using adiabatic demagnetization in the mK range.

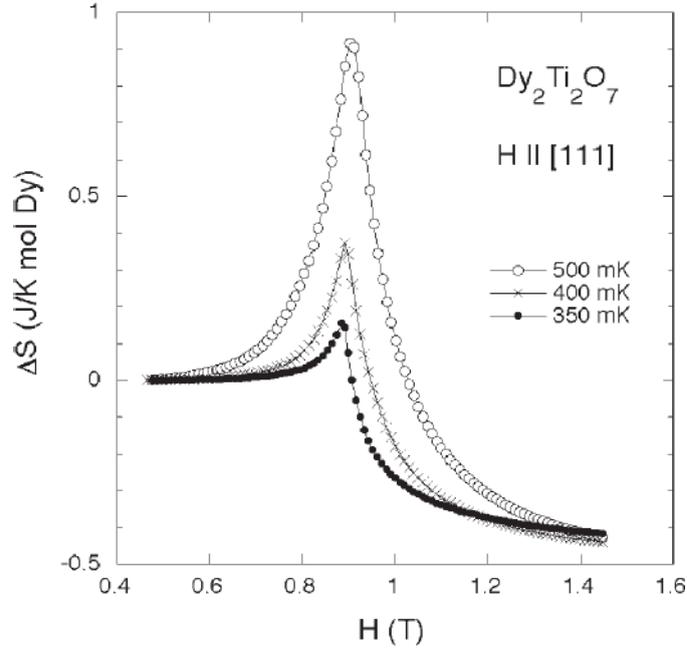

FIG. 9: $\Delta S_M$-H curve measured at 350 mK, 400 mK and 500 K of $Dy_2Ti_2O_7$ sample. Adapted from [53].

Further a large MCE has been observed in the frustrated pyrochlore AFM $Gd_2Ti_2O_7$ single crystal sample reported elsewhere [55]. In $Gd_2Ti_2O_7$, $Gd^{3+}$ ($4f^7$) exhibit orbital momentum L = 0 and only spin value S = 7/2 ground state. Thus, single ion anisotropy and crystal-field splitting become unimportant. Therefore, a excellent classical Heisenberg AFM system with dipole-dipole interactions can be realized in $Gd_2Ti_2O_7$ system. It shows absence of long-range order down to 1 K. The transition below 1 K occurs likely due to dipole-dipole interactions. The values of $\Delta S_M$ and RCP at 4.2 K for demagnetization for 90 kOe field change calculated from specific heat data are 17.89 J/K-mol and 30 J/mol-Gd.

The next material of interest in this series is pyrochlore $Er_2Ti_2O_7$. Here, $Er^{3+}$($4f^{11}$) ion exhibit single ion anisotropy. Therefore, it can be realized as a frustrated XY AFM system on pyrochlore lattice. It exhibits a large value of negative $\theta_{CW}$ = -22 K and displays an AFM transition at ∼ 1.17 K. Figure 10a displays $\Delta S_M$ versus T curve of polycrystalline $Er_2Ti_2O_7$ sample synthesized by solid state route [56]. It increases with reducing the temperature for a field change of 10 T. It shows symmetrical span at ∼ 4.5 K, suggests a second order magnetic phase transition. Figure 10b reveals the value of -$\Delta S_M^{max}$ and RCP enhances as field increases. Moreover, it has shown that the lowest accessible temperature by demagnetizing single crystal of $Er_2Ti_2O_7$ lies down to 100 mK [57]. This property makes it suitable magnetic coolant above 200 mK.

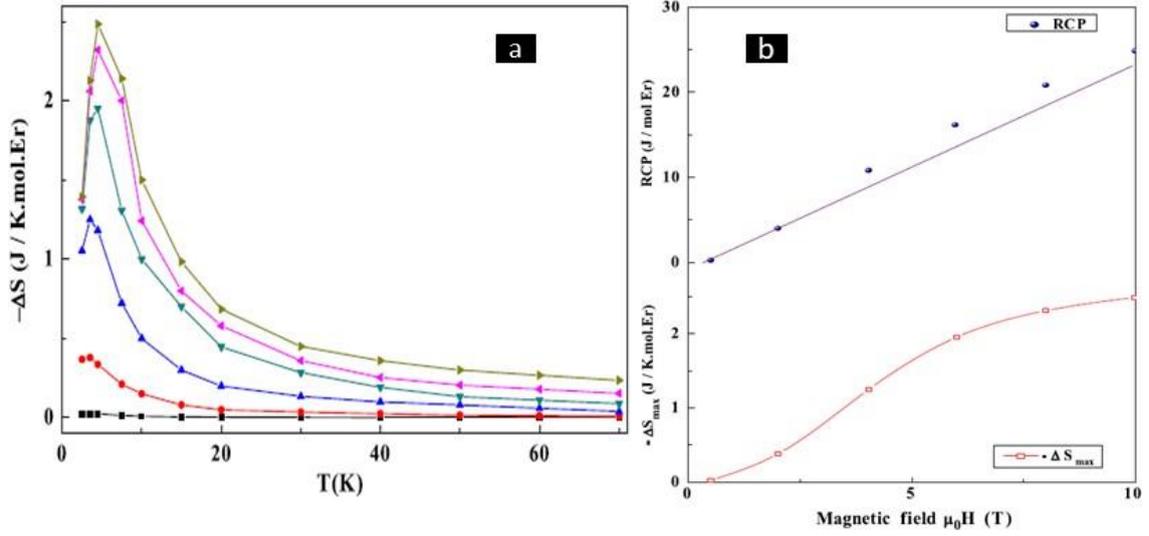

FIG. 10: (a) -$\Delta S_M$ versus T curve at field changes upto 10 T for the sample $Er_2Ti_2O_7$. Adapted from [56].

## C. TiMn based pyrochlore oxides

The 50% replacement of non-magnetic $Ti^{4+}$ with magnetic $Mn^{4+}$ ion in pyrochlore $Er_2Ti_2O_7$ i.e., $Er_2TiMnO_7$ compound makes the system interesting because the interaction between $Mn^{4+}$-($3d^3$) and rare earth ion $Er^{3+}$-($4f^{11}$) may give rise complex magnetic properties. Polycrystalline $Er_2TiMnO_7$ Sample has synthesized by conventional solid-state route and found to be crystallized in monoclinic layered structure with space group P2/M reported elsewhere [58]. The magnetization studies suggest an AFM transition at $T_N$ = 2.05 K with the value $\theta_{CW}$ = -20.29 K. The estimated maximum value of $\Delta S_M^{max}$ found to be 1.7 J/K-mol for the 10 T field change.

Further, the magnetization and MCE has been investigated in cubic pyrochlore systems $Dy_2TiMnO_7$ and $Ho_2TiMnO_7$ synthesized by solid state reaction method [59]. Both compound show FM transition at $T_C$ = 7.6 K ($Dy_2TiMnO_7$) and 7.5 K ($Ho_2TiMnO_7$). System display FM exchange interaction with the value of $\theta_{CW}$ = +10.3 K ($Dy_2TiMnO_7$) and $\theta_{CW}$ = +9.5 K ($Ho_2TiMnO_7$). Figure 11 shows -$\Delta S_M$ versus T curves for both compounds. It is clear that -$\Delta S_M$ is positive in the whole temperature regime and exhibit a maximum around $T_C$. The position of maximum shifts monotonically to higher temperature as applied field enhances, suggesting the presence of second order magnetic phase transition. The estimated values of -$\Delta S_M^{max}$ and RCP under a magnetic field change 7 T, respectively found to be (I) for $Dy_2TiMnO_7$ - 12.50 J/K-kg and 530.3 J/kg, (II) for $Ho_2TiMnO_7$ - 13.95 J/K-kg and 573.2 J/kg, respectively.

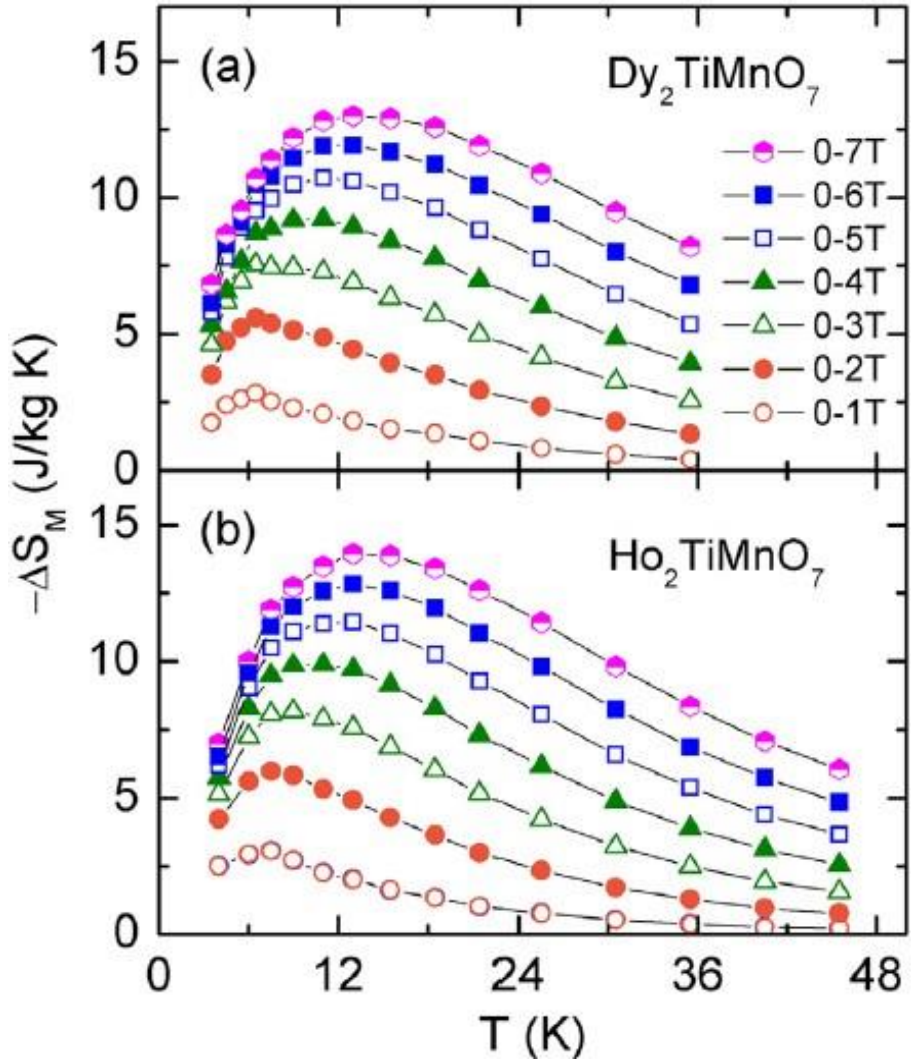

FIG. 11: $-\Delta S_M$ versus T curve at field changes from 0 to 1-0-7 T for the compounds (a) $Dy_2TiMnO_7$, and (b) $Ho_2TiMnO_7$. Adapted from [59].

It is *worth* to mention that some compounds adopted cubic pyrochlore phase and few other systems crystallized in monoclinic structure. Since the available literature are not enough to give a general idea about the parameters which drive the compounds to exist either in pyrochlore or monoclinic structure. As a result, more experimental and theoretical information on various compounds needs to be investigated to get an idea about the relationship between cubic pyrochlore oxide ($Fd\bar{3}m$) and monoclinic layered perovskite (P2/M). In crystal structure of pyrochlore oxide with formula $A_2B_2O_7$, the heavier cation A exhibit eight coordinated while the lighter cations B re-mains in six coordination. On the other hand, in monoclonic compound $Al_2Mn_2O_7$ and pyrochlore sample $Ga_2Mn_2O_7$, although Al and Ga has 3+ valence state, it possess six coordination number, i.e. *interesting* for the formation of pyrochlore structure. For $Sn_2Mn_2O_7$ system, Sn and Mn exhibit 4+ and 3+ oxidation state, respectively. $Sn^{4+}$ may remain in either six or eight coordinated and $Mn^{4+}$ is in six coordinated state. Suppose, if $Mn^{3+}$ goes to $A^{3+}$-site, it should exhibit eight coordination number which might be not *possible* because $Mn^{3+}$ may stay only in six coordinated state.

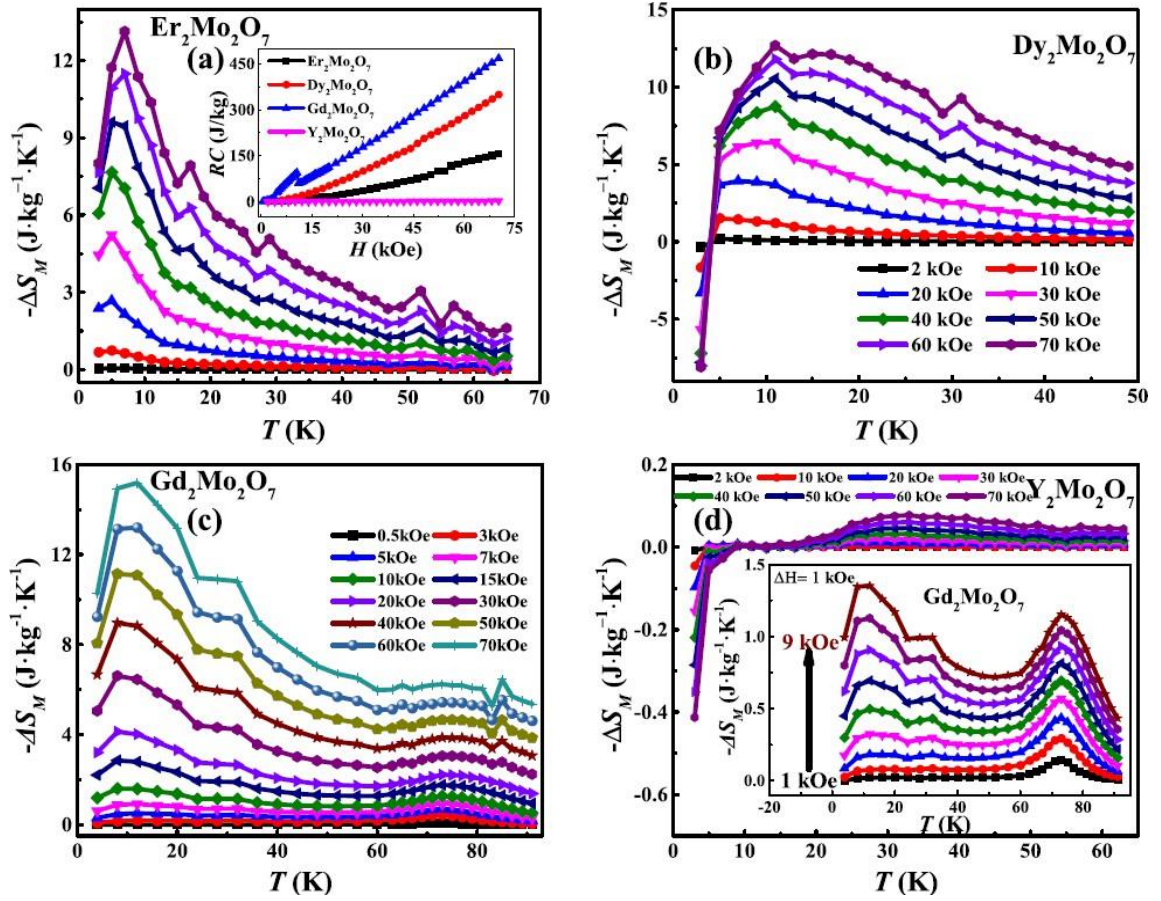

FIG. 12: $-\Delta S_M$ versus T curve for the field change of 7 T for the compounds (a) $Er_2Mo_2O_7$ (b) $Dy_2Mo_2O_7$, (c) $Gd_2Mo_2O_7$ and (d) $Y_2Mo_2O_7$. Inset shows RC versus T curves and inset of (d) displays $-\Delta S_M$ versus T curve for $Gd_2Mo_2O_7$ at low field changes. Adapted from [61].

### D. Mo based pyrochlore oxides

Mo based pyrochlore oxides have attracted due to the various exotic electronic and magnetic ground states with the combinations of rare earth elements [60]. $R_2Mo_2O_7$ series show FM metallic transition for R = Gd, Sm, Nd and spin-glass insulating state for R = Er, Dy, Y, Yb, Tb. $Gd_2Mo_2O_7$ shows long-range FM transition at $T_C$ = 78 K and exhibit reentrant spin-glass state $\sim$ 34 K arises because of AFM exchange interaction between moments of $Mo^{4+}$ and $Gd^{3+}$ ions. On the other hand, $Er_2Mo_2O_7$, $Dy_2Mo_2O_7$ and $Y_2Mo_2O_7$ exhibit only spin-glass state at $\sim$ 20 K [61]. Figure 12 display the -$\Delta S_M$ versus T curve for the series $R_2Mo_2O_7$ (Er, Dy, Gd, Y). A maximum can be seen in -$\Delta S_M$ versus T curves at $\sim$ 10 K for the compound R = Er, Dy and Gd likely due to the ordering of magnetic moments reside on the rare earth ions. The obtained values of -$\Delta S_M$ and RC are found to be; (I) 13.1 J/K-kg, 157.8 J/kg for $Er_2Mo_2O_7$, (II) 12.7 J/K-kg, 349.2 J/kg, for $Dy_2Mo_2O_7$, and (c) 15.2 J/K-kg, 467.9 J/kg for $Gd_2Mo_2O_7$. In $Y_2Mo_2O_7$, magnetization appears because of the magnetic moments of $Mo^{4+}$ ($Y^{3+}$ is nonmagnetic) which is two orders smaller as compared to other three systems. Therefore, it shows negligible value of RC for the field change of 7 T. At low field change of 10.5 kOe, especially $Gd_2Mo_2O_7$ shows table like MCE feature and a significant value of RC 94.4 J/kg which is attributed to the two successive magnetic transition at $\sim$ 10 K and $\sim$ 78 K. It makes the system a potential magnetic coolant in the low temperature range for low field.

### E. Sn based pyrochlore oxides

$Pr_2Sn_2O_7$ have been reported to exhibit a breakthrough dynamic spin-ice state at low temperatures with weak FM interaction $\vartheta_{CW}$ = +0.32 K. Due to large ionic radius and small magnetic moment, it is expected to show a strong quantum phenomenon. Magnetic studies suggested a non-Kramers doublet with large Ising-like single-ion anisotropy and very slow dynamics for the $Pr^{3+}$ (J = 4) crystal field ground states in $Pr_2Sn_2O_7$ pyrochlore. Polycrystalline $Pr_2Sn_2O_7$ sample can be synthesized by conventional solid-state route. The details can be found elsewhere [62]. Figure 13 reveals -$\Delta S_M$ versus T curves shows a broad maximum for field change of higher than 4 T between 2 and 3 K suggest the suppression of MCE by the single-ion anisotropy. It is clear that -$\Delta S_M$ does not favours an enhancement in MCE upto 15 K despite a strong level of magnetic frustration and change of the applied field is of the order of Tesla. Such behaviour may likely due to various orientations of axex of quantization with respect to the applied field direction. Further, the anlaysis using the approximation of a single tetrahedron in a pyrochlore lattice without and with an exchange coupling favours the absence of increased MCE in a dynamic spin-ice system.

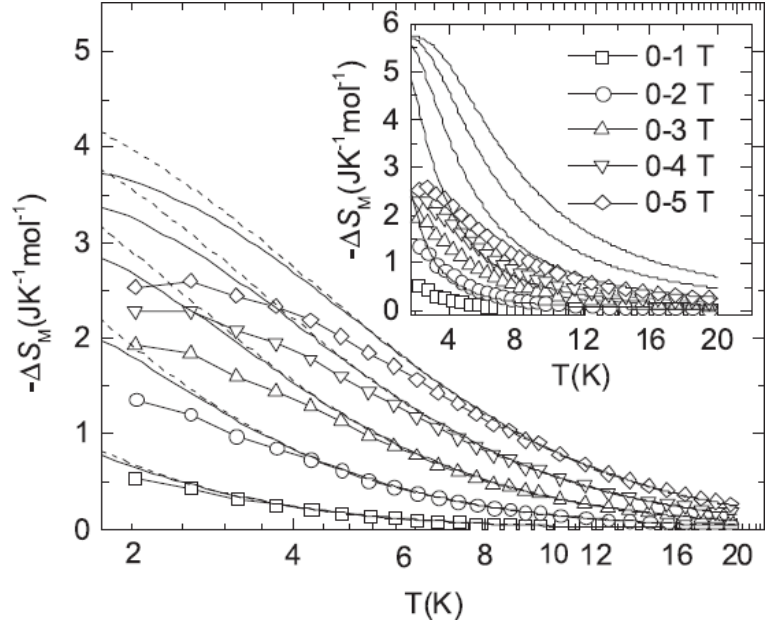

FIG. 13: -$\Delta S_M$ as a function of T of field changes from 5 T for the compound Pr$_2$Sn$_2$O$_7$. The solid and dashed lines represent the theoretical calculations for the involvement of one tetrahedron exhibiting four S = 1/2 Ising spins with and without J/k$_B$, respectively; J = exchange coupling and k$_B$ = Boltzmann constant. Inset shows same data compared with the calculation for an S = 1/2 paramagnet with isotropic g = 5.17. Adapted from [62].

### F. Ir based pyrochlore oxides A$_2$Ir$_2$O$_7$

It have been predicted that the interplay between three comparable energies i.e., electronic correlation, spin-orbit coupling, and crystal field effect combined with magnetic frustration in pyrochlore iridates A$_2$Ir$_2$O$_7$ (A = rare earth element, Y, Bi) leads to exotic electronic and magnetic phases [63–69]. Therefore, it provides a vast template to explore the complex magnetic phases using the MCE as well. The systematic study of A$_2$Ir$_2$O$_7$ shows the enhancement of conductivity with increasing A-site element ionic size. For heavier rare earth elements, R = Pr, Pr$_2$Ir$_2$O$_7$ shows non-magnetic metallic behaviour. While, an antiferromagnetic insulating-like feature for smaller R = Lu, Yb, Er, Ho, Dy, Tb, and Y was observed. The intermediate candidates R = Eu, Sm, and

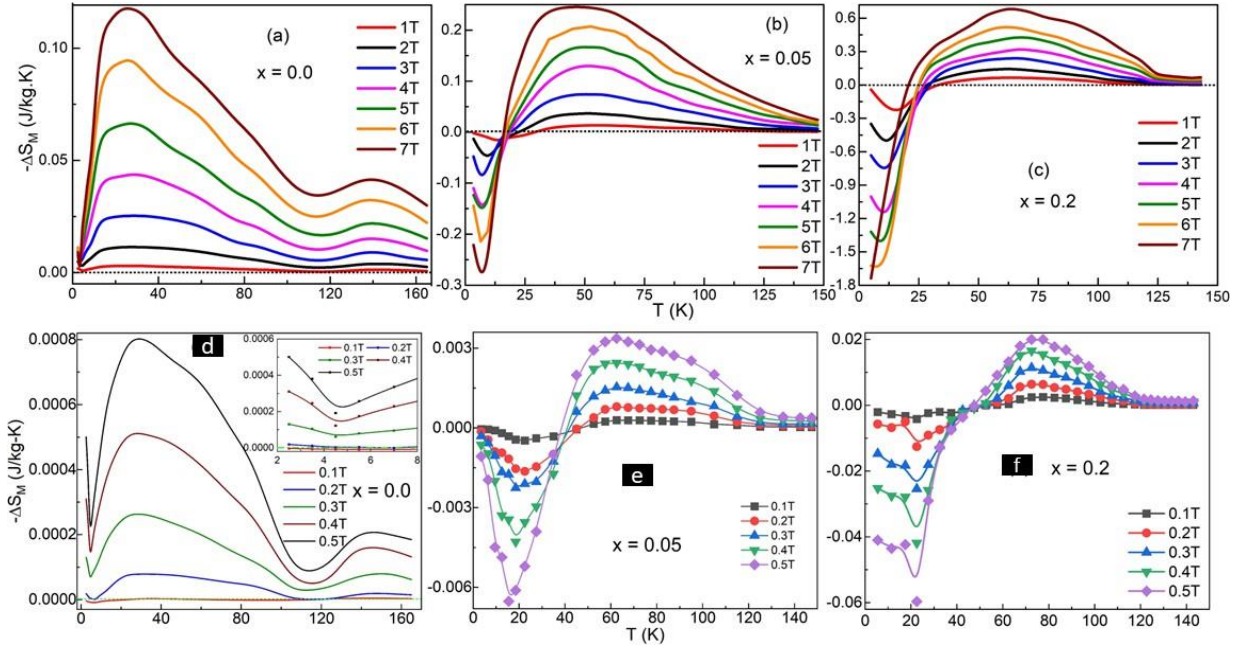

FIG. 14: $-\Delta S_M$ as a function of T of the field change 7 T for the samples (a) x = 0.0, (b) x = 0.05 and (c) x = 0.2. Same curves at low field (d) x = 0.0; inset shows enlarged view of $-\Delta S_M(T)$ at low temperature, (e) x = 0.05 and (f) x = 0.2. Adapted from [70].

Nd shows metal insulator transition, associated with the AFM mean field interaction. $Y_2Ir_2O_7$ has been predicted theoretically to be a magnetic Weyl semimetal with an all-in/all out AFM ground state [63]. Further, MCE have studied in Cr doped $Y_2Ir_2O_7$, i.e., $Y_2Ir_{2-x}Cr_xO_7$ (x = 0.0, 0.05, 0.1 and 0.2) samples [70] and $Er_2Ir_2O_7$ compound [71]. Geometrically frustrated AFM $Y_2Ir_{2-x}Cr_xO_7$ (YICO) series have been reported to exhibit a Griffiths phase (GP)-like state along with cluster-glass-like phase [69]. Ising-like interaction of spins in GP region have also been suggested which provides a hope to achieve high MCE in such frustrated system as it predicted by theory [26, 27]. Pyrochlore cubic $Y_2Ir_{2-x}Cr_xO_7$ series samples can be synthesized by solid state reaction route. Figure 14a-c show $-\Delta S_M$-T curves for various change of magnetic fields upto 7 T [for low field see Fig. 14d-f]. A positive maximum around $T_C$ can be seen for a moderate value of $-\Delta S_M$ for x = 0.0 sample. Surprisingly, $-\Delta S_M$-T becomes negative below $T_N$ for small magnetic field [Fig. 14d]. It turns its sign to positive and reach to a maximum at $T_N$ as field enhances, although the shape of MCE remains same as it is for the low field. Such behaviour arises likely due to spin-flop like transition. Inset of Fig. 14d shows a minimum in $-\Delta S_M$-T curve at $\sim$ 5 K, which might be attributed to Ir spin reorientation. Cr doped samples display the coexistence of conventional

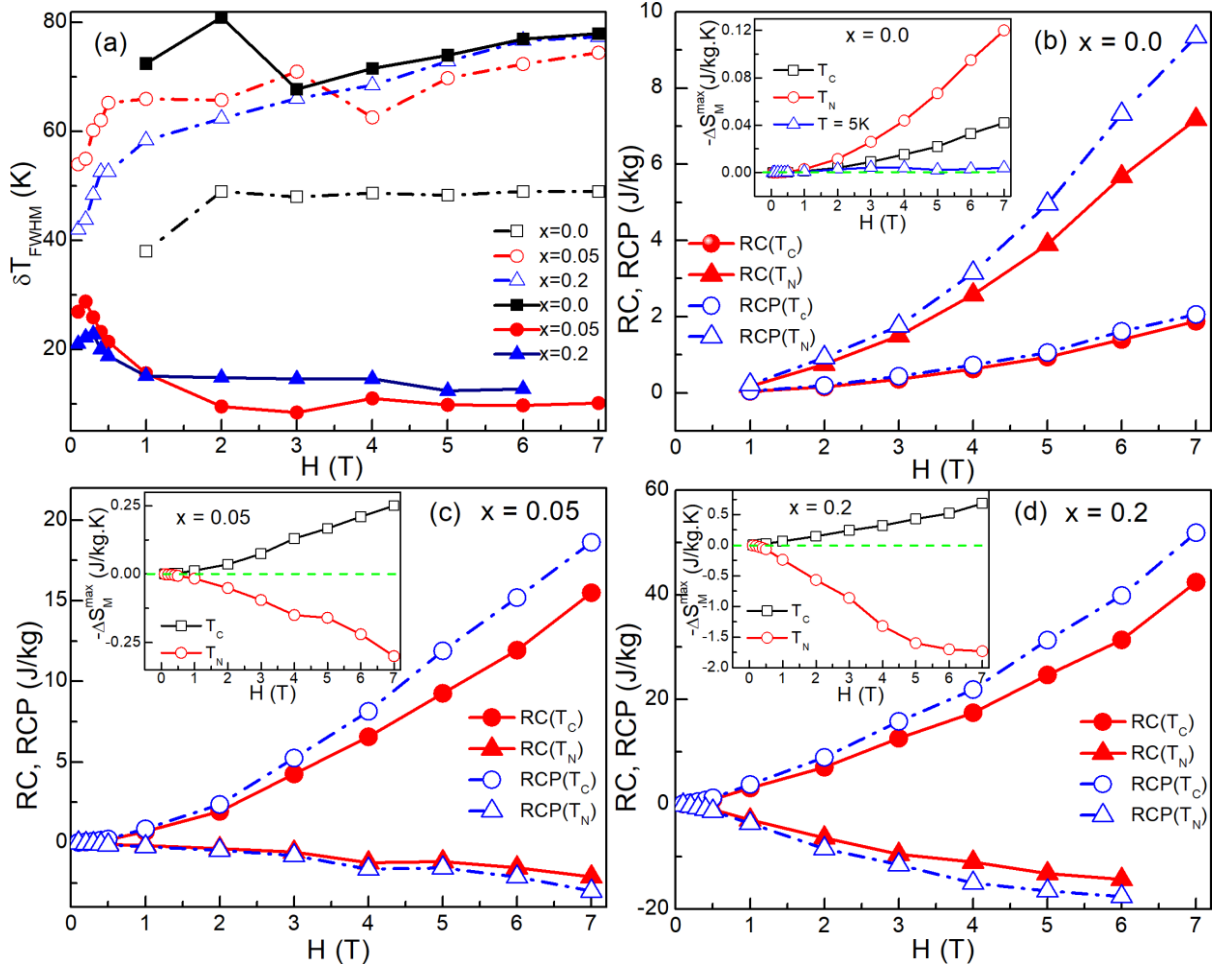

FIG. 15: (a) $\delta_{FWHM}$ versus field H near $T_C$ (hollow symbols) and $T_N$ (solid symbols). Field dependence of RC and RCP for the samples (b) x = 0.0, (c) x = 0.05, and (d) x = 0.2. The inset shows $-\Delta S_M^{max}$ versus H curves. Adapted from [70].

MCE and IMCE for high and as well as low field changes [Fig. 14b,c,e,f]. This crossover from MCE at high temperature to IMCE at low temperature emerges due to the competition between FM and AFM phases likely due to the coexistence of $Ir^{4+}$ and $Ir^{5+}$. YICO series show significant enhancement in the value of $-\Delta S_M$ with Cr substitution, although the value remains smaller than the standard MCE material Gd. RC depends on either a large value of $-\Delta S_M$ or a broad $\delta T_{FWHM}$ of $-\Delta S_M$-T curve or both. Figure 15a shows $\delta T_{FWHM}$ as a function of field H which compensate the small value of MCE, gives rise to a comparable value of RC to the standard MCE materials. The estimated values of RC and RCP around $T_N$ and $T_C$ are displayed in Fig 15b-d. The important outcomes of YICO series are both the 'high temperatures positive' maximum (near $T_C$) and 'low temperature negative' minimum (near $T_N$) in $-\Delta S_M$-T curve extended over a broad range of temperatures. Such coexistence of MCE and IMCE is completely *absent* in the reported pyrochlore oxides. Therefore, $Y_2Ir_{2-x}Cr_xO_7$ series makes a place in the category of new promising magnetic coolant candidates.

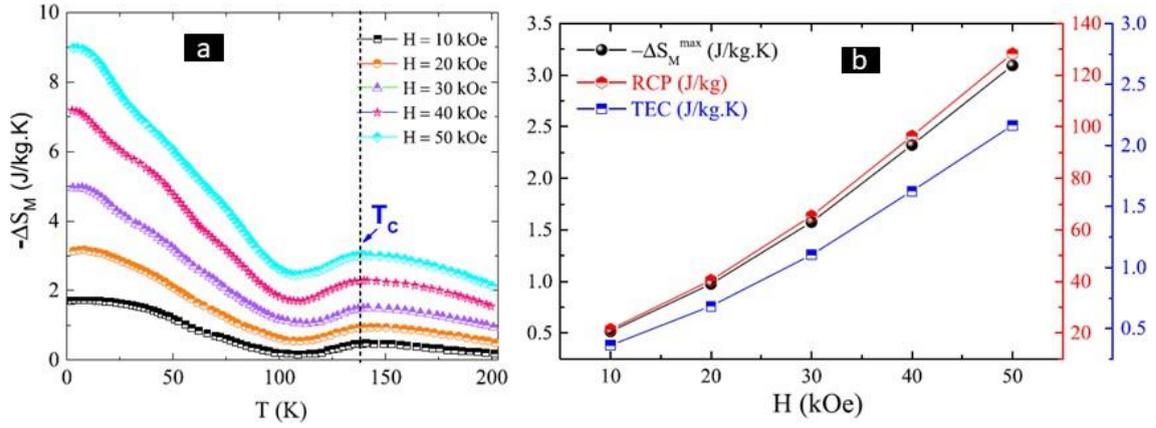

FIG. 16: (a) $-\Delta S_M$ as a function of T under various field changes upto 5 T for $Er_2Ir_2O_7$ compound. (b) Field dependence of $-\Delta S_M^{max}$, RCP and temperature entropy change (TEC). Adapted from [71].

Moreover, the magnetic properties reported in $Er_2Ir_2O_7$ pyrochlore iridate have shown interesting properties due to the interaction between rare earth element $Er^{3+}$ ($4f^{11}$, S = 3/2, L = 6, J = 15/2) and $Ir^{4+}$ ($5d^5$, S = 1/2, L = 0). $Er_2Ir_2O_7$ shows short range ordering below 2 K for Er-sublattice due to frustration and Er-Ir sublattices drive a weak FM transition at $T_C$ = 140 K [72]. Figure 16a shows a broad maximum in $-\Delta S_M$-T curve at $T_C$ = 140 K. The maximum obtained value for $-\Delta S_M^{max}$, and RCP is 3.09 J/K-kg, 128 J/kg, respectively at field change of 5 T [Fig. 16b]. It can be noticed that the value of MCE at $T_C$ is larger than what reported for $Y_2Ir_2O_7$ at 5 T field change. It is likely due to the strong magnetization attributed to the interaction between $Er^{3+}$ and $Ir^{4+}$ ion.

Eventaully, the important results and parameters are summarized in Table I.

## IV. CONCLUSIONS

In conclusions, the magnetocaloric effect and magnetic properties of pyrochlore oxides based on Mn, Ti, TiMn, Sn, Mo and Ir materials are reviewed in detail. The crystal structure, mag- netic frustration, calculation of MCE parameters are also discussed. The Mn based pyrochlores is reported with various A-site ionic radii with the formula $A_2Mn_2O_7$ (A = La, Dy, Ho, Er, Yb,

TABLE I: Important parameters related to the magnetocaloric effect. $(RC)_{T_C}$ and $(RC)_{T_N}$ are the values of Relative Cooling corresponding to $T_C$ and $T_N$, respectively at field change of 5 T. Curie-Weiss temperature is shown by $\theta_{CW}$.

| Sample | Structure | $T_C$ (K) | $T_N$ (K) | $\theta_{CW}$ (K) | $(-\Delta S_M^{max})_{T_C}$ (J/K-kg) | $(-\Delta S_M^{max})_{T_N}$ (J/K-kg) | $(RC)_{T_C}$ (J/kg) | $(RC)_{T_N}$ (J/kg) | References |
|---|---|---|---|---|---|---|---|---|---|
| $Er_2Mn_2O_7$ | Monoclinic | - | 5.47 | -13.92 | - | 2.37 | - | - | [42] |
| $Er_2Mn_2O_7$ | Pyrochlore | 35 | - | +37.4 | 16.1 | - | 522 | - | [45] |
| $Dy_2Mn_2O_7$ | Pyrochlore | 39 | - | +41.5 | 15.2 | - | 515 | - | [46] |
| $Ho_2Mn_2O_7$ | Pyrochlore | 38 | - | +39.8 | 14.0 | - | 420 | - | [46] |
| $Yb_2Mn_2O_7$ | Pyrochlore | 38 | - | +43.5 | 11.5 | - | 300 | - | [46] |
| $La_2Mn_2O_7$ | Monoclinic | 167 | - | +159.0 | 0.65 | - | 31.5 | - | [48, 50] |
| $Fe_2Mn_2O_7$ | Monoclinic | 68 | - | +62.0 | 1.0 | - | 47.25 | - | [47, 50] |
| $Al_2Mn_2O_7$ | Monoclinic | 35 | - | +37.0 | 4.5 | - | 50.25 | - | [49, 50] |
| $Ga_2Mn_2O_7$ | Pyrochlore | - | 42 | -187.87 | - | 1.36 | - | 32.22 | [51] |
| $Ga_2Mn_{1.5}Cr_{0.5}O_7$ | Pyrochlore | - | 41.5 | -151.59 | - | 1.47 | - | 35.79 | [51] |
| $Sn_2Mn_2O_7$ | Monoclinic | 41 | - | +35.97 | 2.94 | - | 30.7 | - | [52] |
| $Er_2Ti_2O_7$ | Pyrochlore | - | 1.17 | -21.54 | - | 2.77 | - | 16.6 | [56] |
| $Er_2TiMnO_7$ | Monoclinic | - | 2.05 | -20.29 | - | 2.28 | - | - | [58] |
| $Dy_2TiMnO_7$ | Pyrochlore | 7.6 | - | +10.3 | 10.73 | - | 273.9 | - | [59] |
| $Ho_2TiMnO_7$ | Pyrochlore | 7.5 | - | +9.5 | 11.45 | - | 267.75 | - | [59] |
| $Er_2Mo_2O_7$ | Pyrochlore | - | 18 | -12.0 | - | 9.6 | - | 80.0 | [61] |
| $Dy_2Mo_2O_7$ | Pyrochlore | 23.0 | - | +4.4 | 10.5 | - | 230.0 | - | [61] |
| $Gd_2Mo_2O_7$ | Pyrochlore | 78.0 | - | +66.3 | 11.1 | - | 340.0 | - | [61] |
| $Y_2Mo_2O_7$ | Pyrochlore | - | 22.0 | -41.5 | - | - | - | - | [61] |
| $Pr_2Sn_2O_7$ | Pyrochlore | - | - | +0.32 | 4.12 | - | - | - | [62] |
| $Er_2Ir_2O_7$ | Pyrochlore | 140.0 | - | -26.0 | 3.09 | - | 128 | - | [71] |
| $Y_2Ir_2O_7$ | Pyrochlore | 130.0 | 15 | -138.0 | 0.022 | 0.067 | 0.94 | 3.9 | [70] |
| $Y_2Ir_{1.8}Cr_{0.2}O_7$ | Pyrochlore | 70.0 | 35.0 | -140.0 | 0.43 | 1.6 | 24.7 | 13.2 | [70] |

Sn, Fe, Ga, and Al). $Er_2Mn_2O_7$ crystallizes in two different crystal structures depending on the synthesis process i.e., monoclinic layered structure with space group P2/M at ambient pressure and cubic pyrochlore structure with space group $Fd\bar{3}m$ at high pressure. Monoclinic $Er_2Mn_2O_7$ sample undergo an antiferromagnetic (AFM) transition with much lower value of RC (2.37 J/kg), while the $Er_2Mn_2O_7$ sample crystallizes in cubic pyrochlore phase shows ferromagnetic (FM) transition exhibiting higher RC value (522 J/kg) than former for the field change 5 T. Such large reversible MCE in cubic pyrochlore $Er_2Mn_2O_7$ emerges due to two sublattice cooperative FM ordering. Pyrochlore $A_2Mn_2O_7$ series display almost similar MCE behabiour with lower value of RC for $Dy_2Mn_2O_7$, $Ho_2Mn_2O_7$ and $Yb_2Mn_2O_7$ samples as compared to cubic pyrochlore $Er_2Mn_2O_7$. A strong single-ion anisotropy does not allow the full alignment in FM fashion, although Dy, Ho, and Yb exhibit larger magnetic moments than Er. Despite the moderate values of RC for the samples crystallize in monoclinic layered perovskite structurei.e., $La_2Mn_2O_7$, $Fe_2Mn_2O_7$ and $Al_2Mn_2O_7$, the MCE values and strength of Griffiths-like state enhances with reduction of A-site ionic radius. In cubic pyrochlore $Ga_2Mn_2O_7$ sample doped with magnetic Cr at magnetic Mn-site, i.e., $Ga_2Mn_{2-x}Cr_xO_7$ (x = 0, 0.1, 0.3, 0.5), the value of RC shows enhancement with Cr doping likely due to mixed oxidation states of $Mn^{3+}$ and $Mn^{4+}$. The series also shows the presence of Griffiths-like state along with glassy phase. The $Sn_2Mn_2O_7$ compound crystallizes in monoclinic layered perovskite structure undergoes FM transition exhibiting moderate RC value. XPS study has suggested 4+ and 3+ valence state of Sn and Mn, respectively, which is *surprising*.

The titanium-based sample $Er_2Ti_2O_7$ crystallized in cubic pyrochlore structure shows AFM transition with low value of RC likely due to single-ion anisotropy which prevents long range ordering. The replacement of 50% nonmagnetic Ti atom with magnetic Mn atom i.e., $Er_2TiMnO_7$ acquires monoclinic structure exhibit moderate value of reversible MCE just above the boiling temperature of helium. The pyrochlore cubic compounds $Dy_2TiMnO_7$ and $Ho_2TiMnO_7$ shows FM transition exhibit significant enhancement in RC value.

The molybdate series $Er_2Mo_2O_7$, $Dy_2Mo_2O_7$, $Gd_2Mo_2O_7$ and $Y_2Mo_2O_7$ crystallizes in pyrochlore structure show spin-glass transition for the samples A = Er, Dy, Y while it undergoes FM transition and subsequently exhibits a reentrant spin-glass behavior for A = Gd sample. The obtained RC value is significantly larger in $Gd_2Mo_2O_7$ compound while $Y_2Mo_2O_7$ exhibit negligible MCE value as compared to the other members of the series $A_2Mo_2O_7$.

Iridium based compounds exhibit the interplay of spin-orbit coupling, electronic correlations, crystal electric field and magnetic frustration which provide a broad window to explore the com-plex magnetic and electronic states. $Er_2Ir_2O_7$ undergoes a weak ferromagnetic transition likely due to the ordering of $Ir^{4+}$ moments. It exhibits significant value of RC likely arises due to the interaction of rare earth element Er and transition metal Ir.

Eventually, the pyrochlore YICO series undergo various magnetic transitions, which is likely attributed to the frustration induced coexistence of FM and AFM clusters. It give rise to the IMCE at low temperature and conventional MCE at high temperature. In this series, magnetic frustration introduced by chemical substitution giving rise to not only coexistence of inverse MCE and conventional MCE but order of magnitude increment of -$\Delta S_M$. Although the value of RC are not large, the conventional MCE and IMCE span over a giant working temperature range. It can be noticed that coexistence of IMCE and MCE is completely *absent* in all reported pyrochlore oxides except YICO series.

Besides a detail review of all reported litterateurs of MCE in pyrochlore oxides, it is interesting to see that $A_2Mn_2O_7$ series crystallizes in monoclinic layered perovskite structure and cubic pyrochlore phase depending on synthesis process. Therefore, more experiments on various compounds are required to shed the light on this point. Hopefully, this article may open up a future direction to encourage the researchers working in various fields associated to the MCE materials.

---